\documentclass[11pt,a4paper]{article}
\pdfoutput=1

\usepackage{jheppub}
\usepackage{latexsym}
\usepackage{multirow}
\usepackage{color}
\usepackage[usenames,dvipsnames,svgnames,table]{xcolor}

\usepackage{graphicx}
\usepackage{epsfig}  
\usepackage{epsf}    
\usepackage{dcolumn}
\usepackage{bm}
\usepackage{dcolumn}
\usepackage{textcomp}
\usepackage{float}
\usepackage{hypcap}
\usepackage[]{hyperref}
\usepackage{adjustbox}
\usepackage{multirow}
\usepackage[utf8]{inputenc}
\usepackage[T1]{fontenc} 
\usepackage{subfigure}
\usepackage{alphalph}
\usepackage{morefloats}
\usepackage{textcomp}
\usepackage{ulem}

\newcommand{\be}{\begin{equation}}
\newcommand{\ee}{\end{equation}}
\newcommand{\ba}{\begin{eqnarray}}
\newcommand{\ea}{\end{eqnarray}}

\newcommand{\blue}[1]{{\color{blue} #1 }}

\title{Impact of CP violation searches at MOMENT experiment with sterile neutrino} 

\author[a]{\,Kiran Sharma}
\author[a]{,\,Sudhanwa Patra}

\affiliation[a]{Department of Physics, Indian Institute of Technology Bhilai, Raipur-492015, India}

\emailAdd{kirans@iitbhilai.ac.in}
\emailAdd{sudhanwa@iitbhilai.ac.in}

\abstract{
We examine the scope of the MOMENT experiment in the context of CP violation searches with the presence of extra eV scale sterile neutrino. MOMENT is a proposed medium baseline neutrino oscillation experiment using muon beams for neutrinos production, making it advantageous over $\pi_0$ background and other technical difficulties. We work over the first oscillation maxima which matches the peak value of flux with a run time of 5 years for both neutrino and anti-neutrino modes. We perform the bi-probability studies for both 3 and 3+1 flavor mixing schemes. The CP violation sensitivities arising from the fundamental CP phase $\delta_{13}$ and unknown CP phase $\delta_{14}$ are explored at the firm footing. Slight deteriorations are observed in CP violations induced by $\delta_{13}$ as the presence of sterile neutrino is considered.  We also look at the reconstruction of CP violations phases $\delta_{13}$ and $\delta_{14}$ and the MOMENT experiment shows significant capabilities in the precise measurement of $\delta_{13}$ phase.
}

\keywords{Neutrino Oscillation, Medium-baseline, Sterile Neutrino, MOMENT}
\begin{document}
\maketitle

\section{Introduction and Motivation}
\label{introduction}

Neutrino physics has made tremendous progress in the past few years. The discovery of the phenomenon of neutrino oscillations~\cite{Wolfenstein:1977ue,Bilenky:1998dt,Giunti:2006fr,Nunokawa:2007qh,Schwetz:2008er,Pascoli:2013wca,Blennow:2013rca,Bellini:2013wra} has not only answered the solar- neutrino problem~\cite{Mikheyev:1985zog,Bethe:1986ej,Haxton:2012wfz,Maltoni:2015kca} and atmospheric-neutrino problem~\cite{Fogli:2003th,Gonzalez-Garcia:2004pka,Kajita:2016gnc,Kajita:2016cak}, but also it has open up the new doors to look at the beyond standard model physics. At present, as per the standard model, there are three active flavors of neutrinos associated with the electron, muon and, tau lepton. The neutrino oscillations in the presence of three active neutrinos are described by six oscillation parameters: two mass square differences i.e. $\Delta m^2_{21} = m^2_2 - m^2_1$, $\Delta m^2_{31} = m^2_3 - m^2_1$, three neutrino mixing angles $\theta_{12}$, $\theta_{13}$, $\theta_{23}$ and one Dirac CP phase $\delta_{CP}$. The values of the oscillations parameters have been constrained by various neutrino experiments ~\cite{DayaBay:2015ayh,RENO:2015ksa,DoubleChooz:2014kuw,DayaBay:2013yxg,Nunokawa:2005nx,deGouvea:2005hk,MINOS:2013utc,Fogli:1996pv}, the only unknown oscillation parameters left to be known precisely are the sign of atmosphere mass -squared difference i.e. $\Delta m^2_{31}$ which will resolve the issue of neutrino mass ordering, the precise measurement of mixing angle $\theta_{23}$ which will help in establishing the correct octant and the value of fundamental CP phase. The long baseline experiments T2K~\cite{T2K:2019bcf} and NO$\nu$A~\cite{NOvA:2021nfi} have independently reported the value of $\delta_{CP}$ but there is a slight tension between the two values. This discrepancy may be because of the systematic uncertainties or it may be an indication of the new physics arising from the presence of sterile neutrinos~\cite{Abazajian:2012ys,Palazzo:2013me,Gariazzo:2015rra,Giunti:2015jba,Giunti:2015wnd,Giunti:2019aiy,Boser:2019rta,Kopp:2013vaa,Sharma:2022qeo} or non-standard interactions~\cite{Ohlsson:2012kf,Farzan:2017xzy,Falkowski:2019kfn,Bischer:2019ttk,Ge:2018uhz}. Also in~\cite{Rahaman:2022rfp}, non-unitarity and Lorentz invariance violation has been investigated as a potential explanation for resolving the tension.

Certain short baseline anomalies are coming from the reactor and Gallium~\cite{Mention:2011rk,GALLEX:1997lja}, accelerator experiments LSND~\cite{LSND:2001aii} and MiniBooNE~\cite{MiniBooNE:2018esg} which hints towards the existence of a hypothetical fourth sterile mass eigenstate. The presence of such right-handed sterile neutrinos will extend the structure of the standard 3 flavor framework. As a result, the number of neutrino oscillation parameters gets enhanced in the minimal extended $3+1$ framework. There are now three additional mixing angles i.e. $\theta_{14}$, $\theta_{24}$, $\theta_{34}$ and two more CP phases. We have the new mass square differences arising from the mixing among fourth mass eigen state $\nu_4$ with mass eigenstates of three active flavors ($\nu_e$, $\nu_\mu$ and, $\nu_\tau$). The presence of sterile neutrino has been widely explored in the literature (see references~\cite{Agarwalla:2018nlx,Choubey:2017cba,Choubey:2017ppj,Haba:2018klh,Coloma:2017ptb,Agarwalla:2016xxa,Berryman:2015nua}). 

The effect of sterile neutrinos over the standard oscillation parameters have been widely explored in the current and future proposed neutrino oscillation experiments like $\rm{T2HK}$~\cite{Hyper-KamiokandeProto-:2015xww},$\rm {T2HKK}$~\cite{Hyper-Kamiokande:2016srs}, and $\rm{DUNE}$~\cite{DUNE:2015lol,DUNE:2016hlj,DUNE:2016evb,DUNE:2016rla}. An important point to remark is that all the above-stated experimental facilities are using the neutrino beams originating from the pions decay, whereas in the proposed work, we look at the potential of upcoming neutrino oscillation experiment $\rm {MOMENT}$(MuOn-decay MEdium baseline NeuTrino beam facility)~\cite{Cao:2014bea} which \blue{will} observe a neutrino beam produced from decaying muons at relatively low energies. As an advantage one can skip the background effects and other technical difficulties at the experimental level. However, it has a non-negligible background arising from atmospheric neutrinos.
 $\rm {MOMENT}$ experiment also provides access to eight different oscillation channels with a high control of systematic uncertainities.

MOMENT is a proposed medium baseline neutrino oscillation experiment with a baseline of 150 km. A detailed description of the experimental facility is provided in section~\ref{sec:experiment}. The matter effects arising from the interaction of neutrinos with the matter potential as they propagate through it are negligible in comparison to the long baseline neutrino experiments. As a result, the fake CP violation scenarios arising from matter terms can be avoided in the MOMENT. The comparative analysis for the precise measurement of delta CP phase has been made among MOMENT, T2K, NO$\nu$A, T2HK, T2HKK, and DUNE where it has been discussed that MOMENT can significantly improve the bounds over $\delta_{CP}$~\cite{Tang:2019wsv}. Furthermore, the effect of non-standard interactions and their effect over the CP phase has been looked at in references(~\cite{Blennow:2015cmn,Bakhti:2016prn,Tang:2017qen,Tang:2019wsv}). In paper~\cite{Tang:2018rer}, authors investigated the invisible decays of the third neutrino mass eigenstate for MOMENT. The benefits of $\rm {MOMENT}$ like beams with opaque detector are explored in~\cite{Tang:2020xoy}. To the best of our knowledge, the effect of sterile neutrinos in the MOMENT facility is not been studied to a much large extent. In this work, we look at the influence of sterile neutrino over the precise measurement of the fundamental CP phase. This work aims to study the capabilities of the MOMENT experiment and put it into context in the global experimental effort in neutrino physics.

This motivates us to study the physics potential of MOMENT experiment in the presence of a hypothetical right-handed eV scale sterile neutrino. We present a detailed discussion of the
behavior of the 4-flavor transition probabilities and understand the bird's eye view of CP trajectory curves in the standard 3-flavor scheme and extend it to our framework of the 3+1 flavor scheme. We perform a prospective study addressing the CP violation sensitivities and look at the degeneracies among different CP phases.

The paper is organized as follows: In Sec.~\ref{sec:probability}, we give the theoretical understanding of the neutrino and anti-neutrino oscillation probabilities in the presence of a sterile neutrino. In Sec.~\ref{sec:prob}, we discuss the quantitative discussion for neutrino and anti-neutrino appearance probabilities for different values of the sterile CP phase. In Sec.~\ref{sec:biprob}, we extend our discussion for the analytical examination of neutrino oscillations at the bi-probability level. In Sec.~\ref{sec:experiment}, we look at the numerical simulations for event rates for signal and background. In Sec.~\ref{sec:CP}, we focus on the CP violation study and the reconstruction among different CP phases. We summarize and conclude our results in Sec.~\ref{sec:Conclusions}.


\section{Transition probability in the 4-flavor scheme}
\label{sec:probability}
\subsection{Theoretical framework}
The formalism of $3+1$ neutrino oscillation can be understood in terms of 
time dependent Schr\"{o}dinger equation in the mass basis as,
\begin{eqnarray}
 i \frac{\partial \nu_j}{\partial t} &=& H_0 \nu_j,
  \label{mass_basis}
  \end{eqnarray}
with $j=1,2,3,4$ and $H_0$ is defined as the effective Hamiltonian in the mass basis and $\nu_j$ being the neutrino mass eigenstate. 
The effective flavor dependent Hamiltonian for $3+1$ neutrino oscillation including matter effects is given by
\begin{equation}
H_{4\nu} \;=\; 
\underbrace{
U_{4\nu}
\left[ \begin{array}{cccc} 0 & 0 & 0 & 0  \\
                          0 & \Delta m^2_{21}/2E & 0 & 0 \\
                          0 & 0 & \Delta m^2_{31}/2E & 0 \\
                          0 & 0 & 0 & \Delta m^2_{41}/2E  
       \end{array}
\right]
U^\dagger_{4\nu}}_{\displaystyle =H_{\rm vac}}
+\underbrace{
\left[ \begin{array}{cccc} V_{CC} & 0 & 0 & 0 \\
                          0 & 0 & 0 & 0 \\
                          0 & 0 & 0 & 0 \\
                          0 & 0 & 0 & -V_{NC}
       \end{array}
\right]}_{\displaystyle =H_{\rm mat}} \;,
\label{HinMatter}
\end{equation}
In case of $3+1$ scenario, i.e, for three active neutrinos and one sterile neutrino, the mixing matrix $U_{4\nu}$ can be parameterised as 
\begin{eqnarray}
U_{4\nu} &=& R\big(\theta_{34}, \delta_{34} \big) \, R\big(\theta_{24}, 0 \big) \, R\big(\theta_{14}, \delta_{14} \big)
 R\big(\theta_{23}, 0 \big) R\big(\theta_{13}, \delta_{13}\big) R\big(\theta_{12}, 0 \big) \nonumber \\
 &\equiv& R\big(\theta_{34}, \delta_{34} \big) \, R\big(\theta_{24}, 0\big) \, R\big(\theta_{14}, \delta_{14} \big) U_{3\nu}   \nonumber \\
\end{eqnarray}
where $U_{3\nu}= R\big(\theta_{23}, 0 \big) R\big(\theta_{13}, \delta_{13}\big) R\big(\theta_{12}, 0 \big)$ is the standard three flavor neutrino mixing matrix. The mass eigenstates are related to flavor eigenstates as $\nu_j = \big[U_{4\nu}\big]_{j \alpha} \nu_{\alpha}$. 

The $4\times4$ real rotation matrices $R_{ij}$ (here R($\theta_{24}$), R($\theta_{23}$), R($\theta_{12}$)) in the $(i,j)$ plane with $2\times 2$ sub matrices are defined as:
\begin{equation}
R_{ij} =
\begin{pmatrix}
         c_{ij} & s_{ij} \\
       - s_{ij} & c_{ij}  \\
\end{pmatrix}
\end{equation}

while the complex rotation matrices (i.e. R($\theta_{34},\delta_{34}$), R($\theta_{14},\delta_{14}$, and R($\theta_{13},\delta_{13}$))in the $(i,j)$ plane are defined by:
\begin{equation}
\tilde {R_{ij}}=
\begin{pmatrix}
         c_{ij} & s_{ij} \\
        -s_{ij}^* e^{-i\delta_{ij}}  & c_{ij}  \\
\end{pmatrix}
\end{equation}

with $c_{ij} = \cos{\theta_{ij}}$ and $s_{ij} = \sin{\theta_{ij}}$. The parametrization of unitary mixing matrix allows the conversion probability to independent of mixing angle $\theta_{34}$ and the corresponding delta phase $\delta_{34}$.

In the S-matrix formalism, the neutrino flavor changes after propagating  a distance $L$ is defined in terms of an evolution matrix $S$ as
\begin{equation}
 \nu_\alpha \big(L \big) = S_{\alpha \beta} \nu_\beta \big(0 \big)\,.
\end{equation}
The key point is that the evolution matrix $S$ satisfies the same Schr\"{o}dinger equation. After some simplification, the form of evolution matrix can be expressed in term $H_{4\nu}$ as
\begin{eqnarray}
S_{\beta \alpha}& = &
\left[\,
\exp\!\left(-i H_{4 \nu} L\right)
\right]_{\beta\alpha}
\;,
\end{eqnarray}
The final expression of neutrino oscillation probability from $\nu_\alpha$ to $\nu_\beta$ with neutrino energy $E$ and baseline $L$ is expressed in terms of evolution matrix as,
\begin{eqnarray}
P(\nu_\alpha\rightarrow\nu_\beta)
& = &
\bigl|\,S_{\beta\alpha}\,\bigr|^2
\end{eqnarray}

\subsection{Appearance probability $P^{4\nu}_{\mu e}$}
The $3+1$ appearance probability for $\nu_\mu \to \nu_e$ transition has been derived in ref.\cite{Klop:2014ima} and is related to three flavour neutrino transition probability $P^{3\nu}_{\mu e}$ and other interference terms arising from sterile neutrino mixing parameters as follows,
\begin{align}\label{Pnue}
	P^{4\nu}_{\mu e}
	\approx
	&\quad \bigg(1-\mathrm{s}^2_{14}-\mathrm{s}^2_{24}\bigg) P^{3\nu}_{\mu e}
    \nonumber \\
	&+4\mathrm{s}_{14}\mathrm{s}_{24}\mathrm{s}_{13}\mathrm{s}_{23}\sin\Delta_{31}\sin(\Delta_{31}+\delta_{13}-\delta_{14}), \nonumber \\
	&-4\mathrm{s}_{14}\mathrm{s}_{24}\mathrm{c}_{23}\mathrm{s}_{12}\mathrm{c}_{12}\sin(\Delta_{21})\sin\delta_{14}, \nonumber  \\
		&+2\mathrm{s}^2_{14}\mathrm{s}^2_{24}\, .
\end{align}
with $\Delta_{ij}=\frac{\Delta m^2_{ij}L}{4E}$. 
The expression for $P^{3\nu}_{\mu e}$ is sum of three contributions i.e, atmospheric $P^\mathrm{ATM}$, solar $P^\mathrm{SOL}$ and their interference term $P^\mathrm{INT}_\mathrm{I}$. 
Looking at the neutrino oscillation parameters from recent global fit~\cite{deSalas:2020pgw} values, which are displayed later in Table.\ref{table_1} 
\begin{equation}
\begin{array}{ll}
s_{23} \;=\; 0.76 \;,\quad &
c_{23} \;=\; 0.65 \;,\\ 
s_{12} \;=\; 0.56 \;,\quad &
c_{12} \;=\; 0.83 \;,\\ 
s_{13} \;=\; 0.15 \;,\quad &
c_{13} \;=\; 0.99 \;.\\
\end{array}
\label{scorders}
\end{equation}
The sine of the reactor mixing angle $s_{13}$ is treated as a small parameter in comparison to other mixing angles and is taken to be of the order of $O(\varepsilon) \approx 0.15$ while all other sines and cosines are $O(1)$. The other parameter which is the ratio between two mass-square differences can be considered as $|\alpha|
\;=\; \dfrac{\Delta m^2_{21}}{|\Delta m^2_{31}|} 
\;\approx\;0.03 \simeq \varepsilon^2$. Also, parameters having sine of the sterile mixing angles $\sin\theta_{14}$ and  $\sin \theta_{24}$ are considered to be small and are of the order of  $\varepsilon$. However, the other sterile neutrino mixing angle $\sin \theta_{34}$ and the corresponding phase $\delta_{34}$ are taken to be zero as the vacuum probability expression is independent of these contributions. Retaining term up to third order in $\varepsilon$ the appearance probability $P^{4\nu}_{\mu e}$ can be simplified as sum of three contributions,
\begin{eqnarray}
\label{eq:Pme_4nu_3_terms}
P^{4\nu}_{\mu e}  \simeq  P^{\rm{ATM}} + P^{\rm {INT}}_{\rm I}+   P^{\rm {INT}}_{\rm II}\,.
\end{eqnarray}
with individual contributions are given by
\begin{eqnarray}
\label{eq:Pme_atm}
 &\!\! \!\! \!\! \!\! \!\! \!\! \!\!  P^{\rm {ATM}} &\!\! \simeq\,  4 s_{23}^2 s^2_{13}  \sin^2{\Delta}\,,\\
 \label{eq:Pme_int_1}
 &\!\! \!\! \!\! \!\! \!\! \!\! \!\! \!\! P^{\rm {INT}}_{\rm I} &\!\!  \simeq\,   8 s_{13} s_{12} c_{12} s_{23} c_{23} (\alpha \Delta)\sin \Delta \cos({\Delta + \delta_{13}})\,,\\
 \label{eq:Pme_int_2}
 &\!\! \!\! \!\! \!\! \!\! \!\! \!\! \!\! P^{\rm {INT}}_{\rm II} &\!\!  \simeq\,   4 s_{14} s_{24} s_{13} s_{23} \sin\Delta \sin (\Delta + \delta_{13} - \delta_{14})\,,
\end{eqnarray}
where $\Delta \equiv \Delta_{31}$.

The first term coming from the atmospheric oscillating parameter $\Delta_{31}$ is a positive quantity providing the leading order contribution to the transition probability. The sub-leading contribution are arising from the interferences of different frequencies. The term
$P^{\rm {INT}}_{\rm I}$ corresponds to the interference of solar and atmospheric frequencies while the term $P^{\rm {INT}}_{\rm II}$ is connected with the interference of atmospheric and sterile frequencies. One can infer from the expression of $P^{\rm {INT}}_{\rm I}$ and $P^{\rm {INT}}_{\rm II}$ that the interference induced by the presence of sterile neutrino is not proportional to $\Delta$ while the solar-atmospheric interference term is directly related with it. As a result, the numerical simulation carried out for the MOMENT experiment working over the first oscillation maxima has a good performance.

\section{Discussion at probability level}
\label{sec:prob}
The impact of matter effects on the appearance probabilities is marginal as MOMENT is a medium baseline experiment with a baseline of $150$km. Thus, the vacuum probabilities expressions can be used effectively while neglecting the suppressed contributions from the MSW effect. For illustration, let us consider  the transition probability in the presence of matter effects up to leading order as~\cite{Klop:2014ima,Cervera:2000kp,Asano:2011nj,Agarwalla:2013tza},
\begin{equation}
\label{eq:Pme_atm_matt}
P^{\rm {ATM}}_m \simeq  (1+ 2 v) P^{\rm {ATM}}\,,
\end{equation}
with the correction factor,
\begin{equation}
\label{eq:v}\,
v = \frac{V}{k} \equiv \frac{2VE}{\Delta m^2_{31}}\,, \quad \quad \mbox{with, }\quad 
 V = \sqrt 2 G_F N_e\,
\end{equation}
For the MOMENT experiment with baseline $150$km and considering first maxima peak energy $E\approx 0.3$GeV, the correction factor is estimated to be $v=0.048$. The correction factor for NO$\nu$A experiment is $v\sim 0.17$ while taking the benchmark value of peak energy $E = 2$ GeV and that is the reason why matter effects are important for long baseline studies. 

\begin{figure}[tbh] 
\includegraphics[scale=0.62]{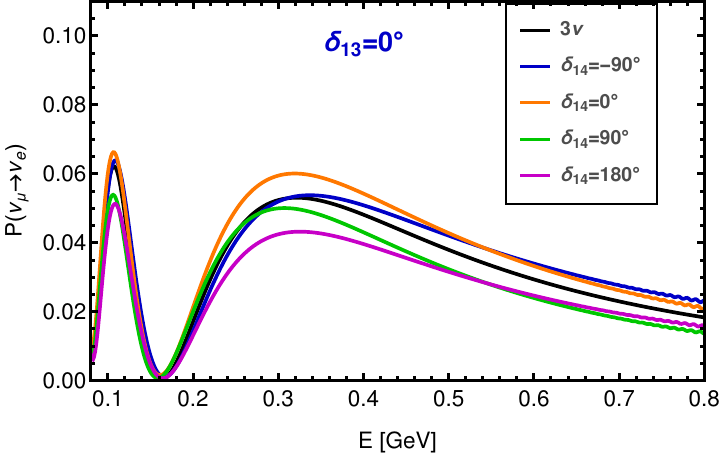}
\includegraphics[scale=0.62]{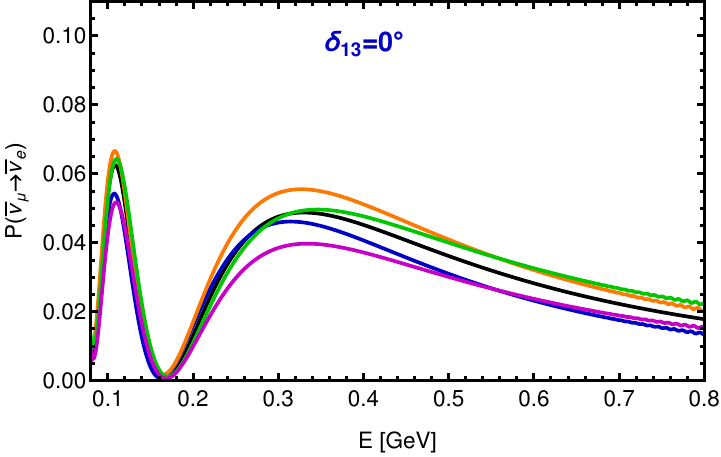}
\includegraphics[scale=0.62]{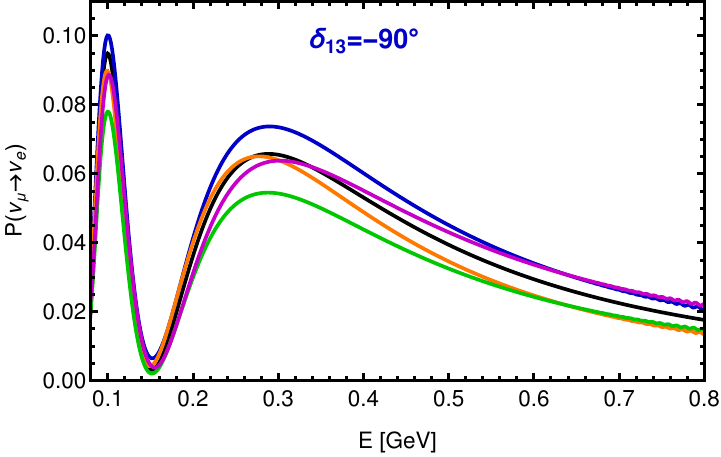}
\includegraphics[scale=0.62]{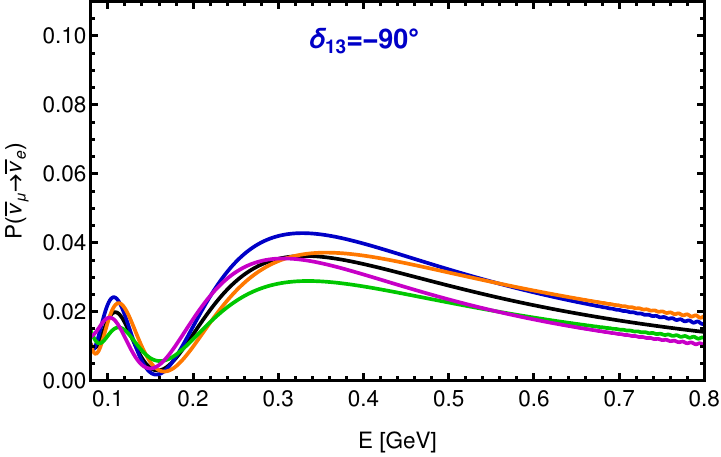}
\caption{Electron appearance probabilities for neutrinos and anti-neutrinos after averaging over the fast oscillations are plotted against energy(varying from 0.1 to 0.8 GeV) with matter density kept fixed at 2.7 g/cc. The oscillation probability for 3 flavor is shown by black curve while the colored curve represents the oscillation probability in 3+1 mixing scenario for four different values of $\delta_{14}$ i.e. $-90^\circ$,$0^\circ$,$90^\circ$, and $180^\circ$. The left column corresponds to the neutrino transition probabilities for two different values of $\delta_{13}$ whereas the right one is for anti-neutrino transition probabilities. }
\label{prob1}
\end{figure}

\begin{figure*}[tbh] 
\includegraphics[scale=0.62]{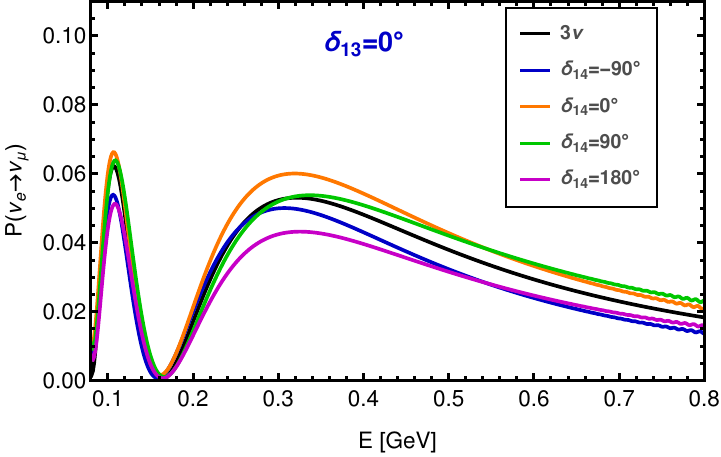}
\includegraphics[scale=0.62]{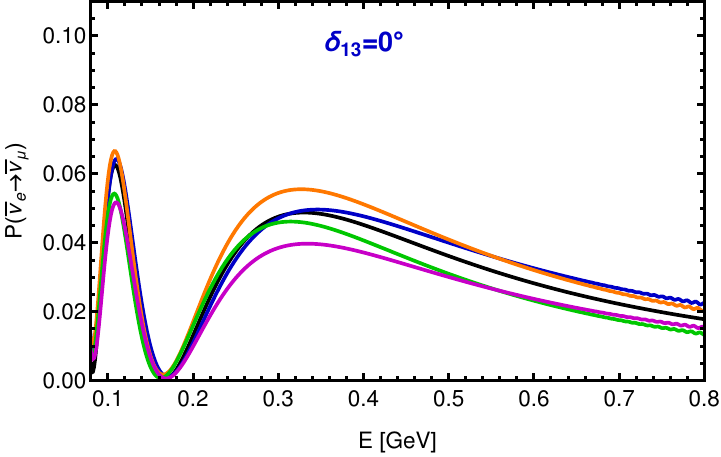}
\includegraphics[scale=0.62]{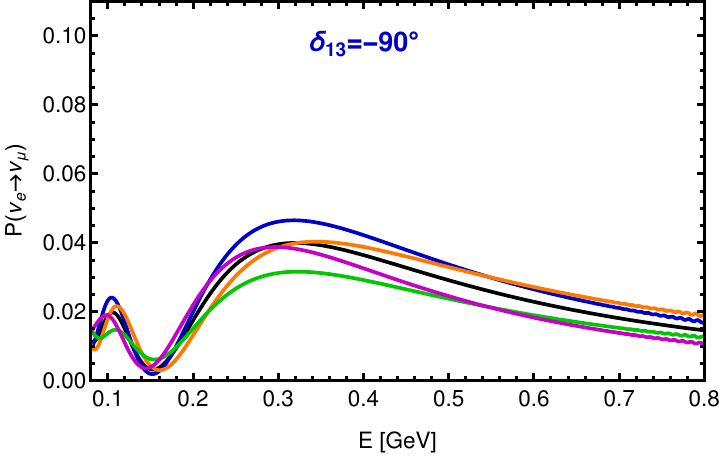}
\includegraphics[scale=0.62]{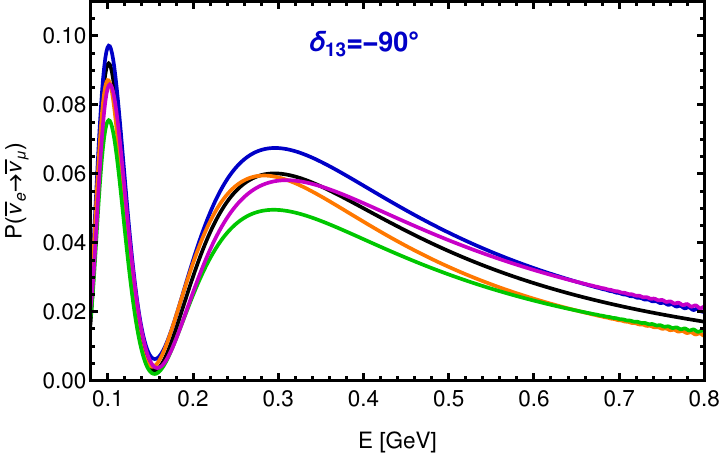}
\caption{Muon appearance probabilities for neutrinos and anti-neutrinos after averaging over the fast oscillations are plotted against energy(varying from 0.1 to 0.8 GeV) with matter density kept fixed at 2.7 g/cc. The oscillation probability for 3 flavor is shown by black curve while the colored curve represents the oscillation probability in 3+1 mixing scenario for four different values of $\delta_{14}$ i.e. $-90^\circ$,$0^\circ$,$90^\circ$, and $180^\circ$. The left column corresponds to the neutrino transition probabilities for two different values of $\delta_{13}$ whereas the right one is for anti-neutrino transition probabilities. }
\label{prob2}
\end{figure*}
We look at the variation of appearance probability channel ($\nu_\mu \rightarrow \nu_e$) and $(\overline{\nu}_\mu \rightarrow \overline{\nu}_e)$ as a function of energy for the fixed value of the fundamental CP phase. The value of $\delta_{13}$ is kept fixed at $0^\circ$ and $-90^\circ$ while the Dirac phase $\delta_{14}$ is allowed to vary over different values as mentioned in the legends in fig~\ref{prob1}. The value of the remaining oscillation parameters has been kept fixed at the benchmark values mentioned in table~\ref{table_1}. The simulations are performed over the constant value of matter density, 2.7 g/cc considered using the Preliminary Reference Earth Model(PREM)~\cite{Dziewonski:1981xy}.
We have taken the value of the sterile mass square difference to be $\Delta m^2_{41} = 1 eV^2$, as a result, the oscillation induced by the sterile presence is averaged out by the finite energy resolution of the detector. We look at the averaged-out behavior of the appearance probabilities. One can also recall that the anti-neutrino appearance probability can be obtained from that of neutrinos by changing the sign of the matter potential and the fundamental and sterile CP phases i.e.
\begin{equation}
P_{\overline{\nu}_{\alpha} \rightarrow \overline{\nu}_{\beta}}(V, \delta_{13}, \delta_{14}) = P_{\nu_{\alpha} \rightarrow \nu_{\beta}}(-V, -\delta_{13}, -\delta_{14}) 
\end{equation}

As evident from eqn~\eqref{eq:Pme_atm_matt}, the leading order contribution to the  appearance  probability will be more for neutrinos with $V>0$ than anti-neutrinos with $V<0$. The solid black line refers to the 3 flavor appearance probability while the colored lines correspond to the different $3+1$ flavor mixing scenario. The blue line indicates the contribution arising from $\delta_{14} = -90^\circ$, orange for $\delta_{14} = 0^\circ$, green for $\delta_{14} = 90^\circ$ and magenta for $\delta_{14} = 180^\circ$. The left column of the figure represents the appearance probability for neutrinos while the right  column explains the behavior of anti-neutrino appearance probability. The curves are peaked around the first oscillation maxima for MOMENT experiment i.e $\approx 0.3 ~GeV$. One can notice that the amplitude and shape of probability are strongly dependent on the value of sterile CP phase $\delta_{14}$. There is a decrease in the amplitude for the anti-neutrino oscillation probability as expected in accordance to eqn~\ref{eq:Pme_atm_matt} and which will be further understood for the event spectrum in section~\ref{sec:experiment}.  Moreover, the plot also shows the mutual swapping for curves as one makes a transition from neutrino to anti-neutrino probability.
\noindent
In fig~\ref{prob2}, we discuss the neutrino and anti-neutrino muon appearance probability channels $(\nu_e \rightarrow \nu_\mu)$ and $(\overline{\nu}_e \rightarrow \overline{\nu}_\mu)$ for two different values of $\delta_{13}$. The electron and muon apperance channels can be obtained from each other using the simple relation,
\begin{equation}
P_{\nu_\alpha \rightarrow \nu_\beta}(\delta_{13},\delta_{14}) = P_{\nu_\beta \rightarrow \nu_\alpha}(-\delta_{13}, -\delta_{14})
\end{equation}
\noindent
We see that oscillation probabilities have strong dependance on $\delta_{14}$ phase and there is  mutual swapping among the curves for different values of $\delta_{14}$. There is a change of an overall normalization in 3+1 flavor mixing case in comparision to the standard 3-flavor case.


\section{Biprobability analysis}
\label{sec:biprob}

The bi-probability curves as the name suggest are the parametric curves that allow us to trace the bi-probability space spanned together by the neutrino and anti-neutrino oscillation probabilities. The idea of bi-probability curves in the 3 flavor scheme was first introduced in the work~\cite{Minakata:2001qm}. The oscillation probability in the 3 flavor scheme is defined as:

\begin{equation}
P = P_0 + A(\cos\Delta \cos\delta_{13} - \sin\Delta \sin\delta_{13})
\label{P_vac}
\end{equation}

\begin{equation}
\overline{P} = \overline{P_0} +\overline{A}(\cos\Delta \cos\delta_{13} + \sin\Delta \sin\delta_{13})
\label{anti_P_vac}
\end{equation}
\noindent
where $A = 8s_{13} s_{12} c_{12} s_{23} c_{23}(\alpha \Delta_{31}) \sin\Delta_{31}$, $\alpha = \Delta m^2_{21}/\Delta m^2_{31} $ , and $P_0 \approx (1+2v)P^{ATM}$ 
\newline
Thus, the bi-probability curves display the effects coming from the fundamental CP phase ($\sin\delta_{13}$ term which is CP violating, $\cos\delta_{13}$ term which is CP conserving), and the matter effects arising from the presence of term A, all in a single diagram. Thus, bi-probability provides us with a bird's eye view important for the mass hierarchy and CP violation studies. Since the MOMENT is a medium baseline facility, it is not worth performing the analysis for mass hierarchy sensitivities. In this work, we fix ourselves to a normal hierarchy and look at the CP trajectory diagrams for neutrino and anti-neutrino oscillation probabilities.  The fundamental CP phase $\delta_{13}$ is varied in the range from $-\pi$ to $\pi$ and biprobability space is spanned. As the probabilities involve the periodic function sine and cosine, as a result, the space spanned must form a closed trajectory as the $\delta_{13}$ is varied. We generalize the bi-probability representation to the  3+1 flavor scheme where the value of $\delta_{14}$ is kept fixed at different values while $\delta_{13}$ is varied over the entire range.

Under the adiabatic approximation, the equation~\eqref{P_vac} and ~\eqref{anti_P_vac}, can be re-expressed as:
\begin{equation}
P = l\cos\delta_{13}+ m\sin\delta_{13}+n
\end{equation}

\begin{equation}
\overline{P} = \overline{l}\cos\delta_{13}- \overline{m}\sin\delta_{13}+\overline{n}
\end{equation}
where
\begin{align*}
l = A \cos\Delta,~ m = -A\sin\Delta,~ n= P_{0} \, \nonumber \\
\overline{l} = \overline{A} \cos\Delta,~ m = -\overline{A}\sin\Delta,~ n= \overline{P}_{0}
\end{align*}

A deeper understanding of the CP-trajectory curves for 3 flavor analysis can be obtained by eliminating the $\delta_{13}$ factor from above equations and under the assumption $A = \overline{A}$ which holds true in vacuum, we obtain the equation followed by neutrino and anti-neutrino probability as:
\begin{equation}
\bigg(\frac{P+\overline{P}-2n}{2l}\bigg)^2 + \bigg(\frac{P-\overline{P}}{2m}\bigg)^2 = 1
\end{equation}

This is the equation of an ellipse where the lengths of the major and the minor axes are measures for the coefficients of $\sin\delta_{13}$ and $\cos\delta_{13}$, respectively, in the oscillation probability. Further, the minor or major axis is always inclined at an angle of $45^\circ$ as visible in bi-probability plots.

The idea of bi-probability plots in 3 flavor scheme can be extended to $3+1$ flavor analysis where we have some additional sources for CP violation arising by sterile neutrino presence as explored in ~\cite{Agarwalla:2016mrc} and the references therein. The analytic expression for the neutrino transition probabilities in case of $3+1$ neutrino oscillation is recast into a compact form as,
\begin{eqnarray}
 P(\nu) \equiv P = P_0 + A \cos \big( \Delta + \delta_{13} \big) 
           + B \sin \big( \Delta - \delta_{14}+ \delta_{13} \big)  \,
 \label{eq:Parametric+nu}
 \end{eqnarray}
where the first term $P_0= P^{\rm {ATM}} \!\! \simeq\,  4 s_{23}^2 s^2_{13}  \sin^2{\Delta}$ is independent of phase factor while the factors which are independent of phase contained in second and third term are, $A=8 s_{13} s_{12} c_{12} s_{23} c_{23} (\alpha \Delta)\sin \Delta$ and $B=4 s_{14} s_{24} s_{13} s_{23} \sin\Delta$. 
After a few simplifications using trigonometric relations, the transition probability given in eq.(\ref{eq:Parametric+nu}) modifies to
\begin{eqnarray}
 P = P_0 + A^\prime \cos\delta_{13} 
           + B^\prime \sin \delta_{13} 
 \label{eq:Parametric+nu+modified}
 \end{eqnarray}
 
 Similarly, the simplified antineutrino transition probability is given by
 \begin{eqnarray}
 \overline{P} = \overline{P}_0 + \overline{A}^\prime \cos\delta_{13} 
           - \overline{B}^\prime \sin \delta_{13} 
 \label{eq:Parametric+nu+modified}
 \end{eqnarray}
 
 where the coefficients $A^\prime$, $B^\prime$,$\overline{A}^\prime$ and $\overline{B}^\prime$ are defined as follows
 \begin{equation}
 A^\prime = A \cos \Delta + B \sin \big(\Delta-\delta_{14} \big)
 \label{A prime}
 \end{equation}
 
 \begin{equation}
 B^\prime = -A \sin \Delta + B \cos \big(\Delta-\delta_{14} \big)
 \label{B prime}
 \end{equation}
 
 \begin{equation}
 \overline{A}^\prime = \overline{A} \cos \Delta + \overline{B} \sin \big(\Delta+\delta_{14} \big)
 \label{A_bar prime}
 \end{equation}
 
 \begin{equation}
 \overline{B}^\prime = -\overline{A} \sin \Delta + \overline{B} \cos \big(\Delta+\delta_{14} \big)
 \label{B_bar prime}
 \end{equation}

The detailed derivation for obtaining the CP-trajectory is mentioned in the appendix~\ref{appendix} obtained by eliminating the $\delta_{13}$. The angle of inclination is obtained by comparing the trajectory equation with the general equation of ellipse and is obtained as:

\begin{eqnarray}
\tan 2\theta = \frac{(B^2 - A^2)\cos 2\Delta - 2AB\sin 2\Delta \cos\delta_{14}}{2 AB\sin\delta_{14}}
\end{eqnarray}

It is clear from the expression that the orientation of the ellipse is strongly dependent on the value of sterile CP phase $\delta_{14}$ and the parameter $\Delta$. The following conclusions can be marked depending on the values of these parameters:
\begin{enumerate}
\item Firstly, under the vanishing condition of factor $\sin \delta_{14}$ (i.e. $\delta_{14} = n\pi$), the inclination becomes $\theta = \pi/4$. The orientation of the elliptical trajectory can be determined by looking at the sign of numerator term. If the numerator is a positive definite quantity the orientation is counter-clockwise while it is clockwise for negative numerator.
\item Also, if $\Delta \approx n\pi/2$, which is true for MOMENT experiment which works in the first oscillation maxima, the inclination angle simplifies as
\begin{equation}
\tan 2\theta \approx \frac{0.366}{\sin \delta_{14}}
\end{equation}
Now there are again two possible inclinations for $\delta_{14}\rightarrow (2n-1)\pi/2$, depending upon the value of n (where $n = {1,2,3...}$). For the positive sign in denominator, the inclination of minor axis is $\approx 10^\circ$ whereas for the negative sign the major axis is inclined by $\approx -10^\circ$.
\end{enumerate}

\begin{figure*}[h]
\includegraphics[scale=0.60]{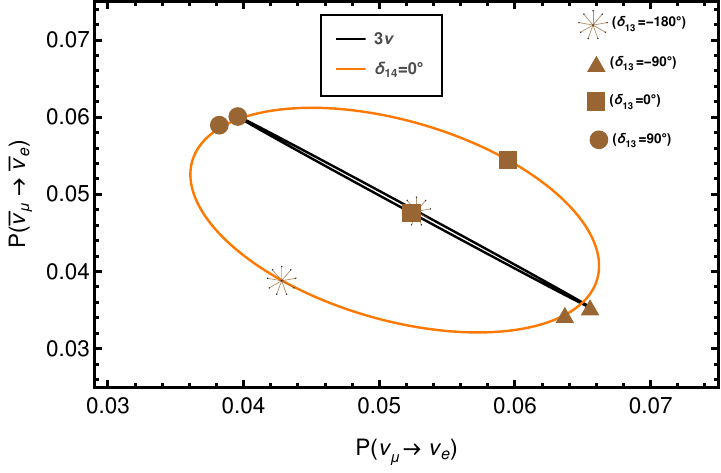}
\includegraphics[scale=0.61]{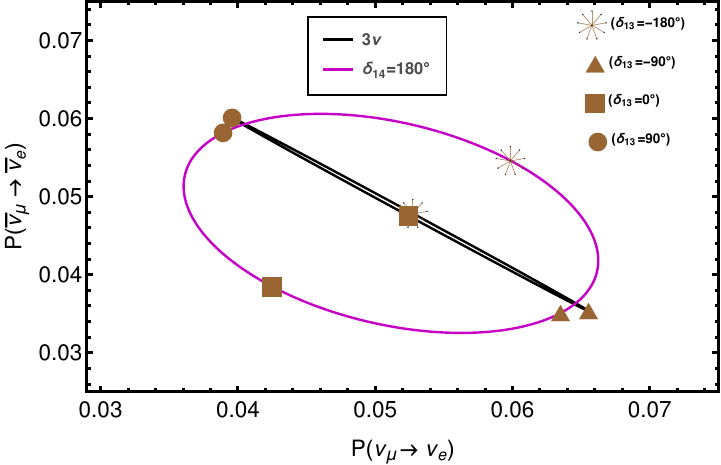}
\includegraphics[scale=0.56]{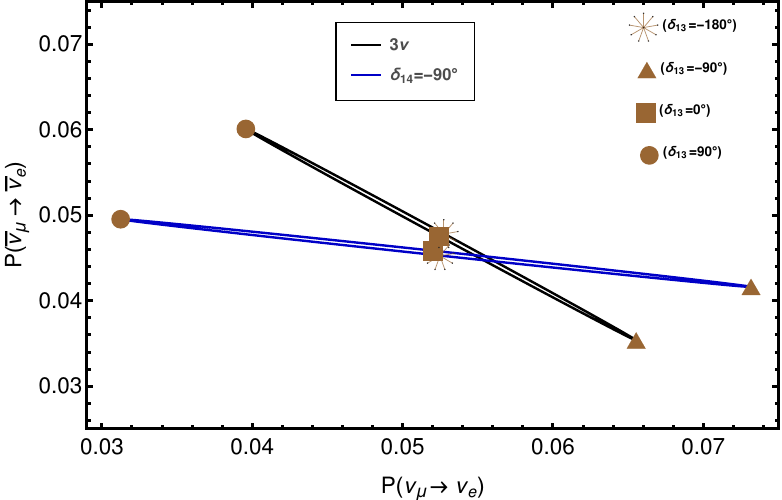}
\includegraphics[scale=0.57]{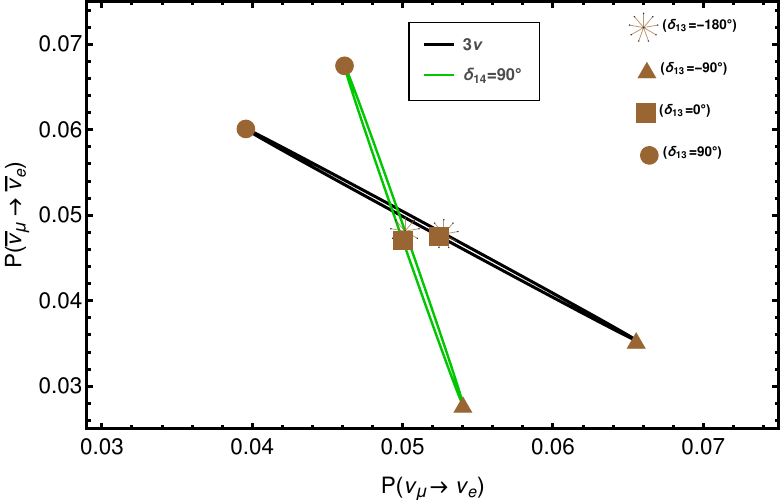}
\caption{Bi-probability plots under the normal hierarchy are shown for both 3 flavor and 3+1 flavor mixing scheme. The CP  trajectory diagram for the 3-flavor is drawn by black color while the orange, magenta, blue and green closed curves corresponds to the fixed values of $\delta_{14}$ phase as mentioned in each legend. The neutrino energy is kept fixed at its first oscillation peak value of 0.3 GeV.  The CP-phase $\delta_{13}$ is varied in the range $[-\pi,\pi]$. Four different values of $\delta_{13}$ phase (i.e. $0^\circ$, $180^\circ$, $-90^\circ$, and $90^\circ$) are marked by different symbols for comparing the orientation of ellipses.}
 \label{bi-prob}
\end{figure*}

The discussion can be easily confirmed from the bi-probability plots in fig~\ref{bi-prob}. We have shown the dependency of ellipse inclination for four different values of $\delta_{14}$. The parameter $\delta_{13}$ is varied over the complete allowed range $[-\pi,\pi]$ while $\delta_{14}$ is kept fixed for each particular plot as mentioned in the legend. The solid black ellipse is drawn for the variation of neutrino and anti-neutrino probabilities in the $3$-flavor scheme. The centers of the ellipses in the $3+1$ flavor scheme are almost coinciding with the centers in the $3$ flavor scheme. The slight variation is arising from the negligible matter effects. We have marked special symbols for highlighting four different fundamental CP phase values i.e $\delta_{13} = {-180^\circ, -90^\circ, 0^\circ,  90^\circ}$. For $\delta_{14} = -180^\circ$ and $\delta_{14} = 0^\circ$, the inclination is $45^\circ$ as seen by the orange and magenta curves. There is a swapping among the values of $\delta_{13}$ phase pointing towards the tracing of elliptical trajectory as expected.
While the blue and green curves plotted under $\delta_{14} = -90^\circ, 90^\circ$ are inclined by $ -10^\circ$ and $ 10^\circ$ respectively.

\section{Experimental Details and Event Spectra}
\label{sec:experiment}
In this section, we shed light on the specifications of the experimental setup MOMENT. As the name suggests MOMENT is a muon-decay medium-baseline neutrino beam facility that uses a continuous-wave proton beam of energy, $15$~MW. Since ideally, it is very difficult for any solid target to withstand such a high power, so mercury jets are adopted as the target. Neutrinos are produced from the muons in a pion decay channel. The $\mu^+$ decays via channel $\mu^+ \rightarrow \bar{\nu}_{\mu} \nu_e e^+$ while $\mu^-$ decays as $\mu^- \rightarrow \nu_{\mu} \bar{\nu}_{e} e^-$. Thus, we have the four neutrino flavors i.e. ${\nu}_{e}$, ${\nu}_{\mu}$, $\bar{\nu}_{e}$ and $\bar{\nu}_{\mu}$. Hence, the MOMENT setup would allow the study of eight oscillation channels (i.e. $\nu_e \rightarrow \nu_e$, $\nu_e \rightarrow \nu_\mu$, $\nu_\mu \rightarrow \nu_e$, $\nu_e \rightarrow \nu_\mu$ as well as their corresponding CP-conjugate partners). To provide sensitivity towards CP-violating, the distinction between the neutrinos and anti-neutrinos is achieved by $500$~kilo ton fiducial Gd-doped Water Cherenkov detector as the baseline detector. The final neutrons captured on Gd provides us a way to distinguish neutrinos from anti-neutrinos with the known interactions of neutrinos with nucleons as $\nu_e + n \rightarrow p + e^-$, $\nu_\mu + n \rightarrow p + \mu^-$, $\bar{\nu}_e + p \rightarrow n + e^+$, $\bar{\nu}_\mu + p \rightarrow n + \mu^+$. The number of the proton on target (POT) with a proton beam power of $15$~MW and with a proton energy of $1.5$~GeV can be calculated as:
\begin{equation}
x = \frac{W \times y \times t \times 10^{16}}{1.6 \times E_{p}}
\end{equation}
\noindent
where $W$ is beam power and $E_{p}$ is the proton energy. The operational time in one year is represented by t and is roughly $\approx 10^{7}$~seconds, while $y$ is number of years protons are deposited on the target. For MOMENT experiment, POT will be $6.2\times 10^{22}$. The fluxes for neutrinos and anti-neutrinos arising from $1.5$~ GeV peaks around $0.030$~GeV while the total energy is varied in range from $0.010-0.80$~GeV. The simulations are performed by assuming the equal run rate for neutrino and anti-neutrino mode. We used an uncertainty $2.5 \%$ on the signal and $5\%$ uncertainty over the background for both neutrinos and anti-neutrino modes. The simulation details for mentioned in table~\ref{table_2}. The neutrino fluxes and cross sections used in simulations are taken from ~\cite{Tang:2020jeh}.

\begin{table}[tbp]
	\centering
	\setlength{\extrarowheight}{0.1cm}
	\begin{tabular}{|c|c|c|c|}
		\hline
      \textbf{Parameters} & \textbf{True Value} & \textbf{Marginalization Range}  & \textbf{3$\sigma$ interval} \\
      \hline
      \hline
      $\sin^2\theta_{12}$ & 0.318 & Not Marginalized & 0.271–0.369 \\
      \hline
      $\sin^2\theta_{13}$ & 0.022 & Not Marginalized & 0.0200–0.02405\\
      \hline
      $\sin^2\theta_{23}$ & 0.574 & [0.38,0.64] & 0.433–0.608\\
      \hline
      $\sin^2\theta_{14}$ & 0.02 & Not Marginalized & 0–0.0996\\
      \hline
      $\sin^2\theta_{24}$ & 0.02 & Not Marginalized & 0–0.0996\\
      \hline
      $\sin^2\theta_{34}$ & 0 & Not Marginalized & 0–0.1894\\
      \hline
      $\delta_{13} $ & [-180,180] & [-180,180] & [-180,180]\\
      \hline
      $\delta_{14} $ & [-180,180] & Not Marginalized & [-180,180]\\
      \hline
      $\delta_{34} $ & 0 & Not Marginalized & [-180,180]\\
      \hline
      $\Delta m^2_{21} (eV^2)$ & $7.50\times 10^{-5}$ & Not Marginalized & $(6.94-8.14)\times 10^{-5}$\\
      \hline
      $\Delta m^2_{31} (eV^2)$ & $2.55\times 10^{-3}$ &  $[2.4,2.7]\times 10^{-3}$ & $(2.47-2.63)\times 10^{-5}$\\
      \hline
      $\Delta m^2_{41} (eV^2)$ & $1$ & Not Marginalized & –\\
      \hline
	\end{tabular}
	\caption{The various parameters and their true value used to simulate the data in GLoBES are mentioned in columns 1 and 2. The standard oscillation mixing parameters are taken from global fit analysis~\cite{deSalas:2020pgw} while 3+1 oscillation parameters are chosen from~\cite{Dentler:2018sju}. The third column depicts the range over which $sin^2 \theta_{23}$ , $\delta_{13}$ , and $\Delta m^2_{31}$ are varied while minimizing the $\chi^2$ to obtain the final results.}	
 \label{table_1}	
\end{table}

\begin{table}[tbp] 
	\centering
	\setlength{\extrarowheight}{0.1cm}
	\begin{tabular}{|c|c|}
		\hline
      \textbf{Characteristics} & \textbf{MOMENT} \\
      \hline
      \hline
      Beam power & 15MW \\
      Fiducial Detector mass  & 500 kton Gd doped Water Cherenkov\\
      Baseline & 150 km\\
      Flux peaks at & 0.3 GeV\\
      First Oscillation maxima  & $\approx 0.3$ GeV\\
      Number of bins & 20 \\
      Bin Width & 0.0395 GeV \\
      Energy Resolution & 0.12/$\sqrt{E}$ \\
      \hline
	\end{tabular}
	\caption{The experimental specifications of MOMENT experiment.}
	\label{table_2}	
	\end{table}

\begin{figure*}[htb!]
\centering
\includegraphics[scale=0.45]{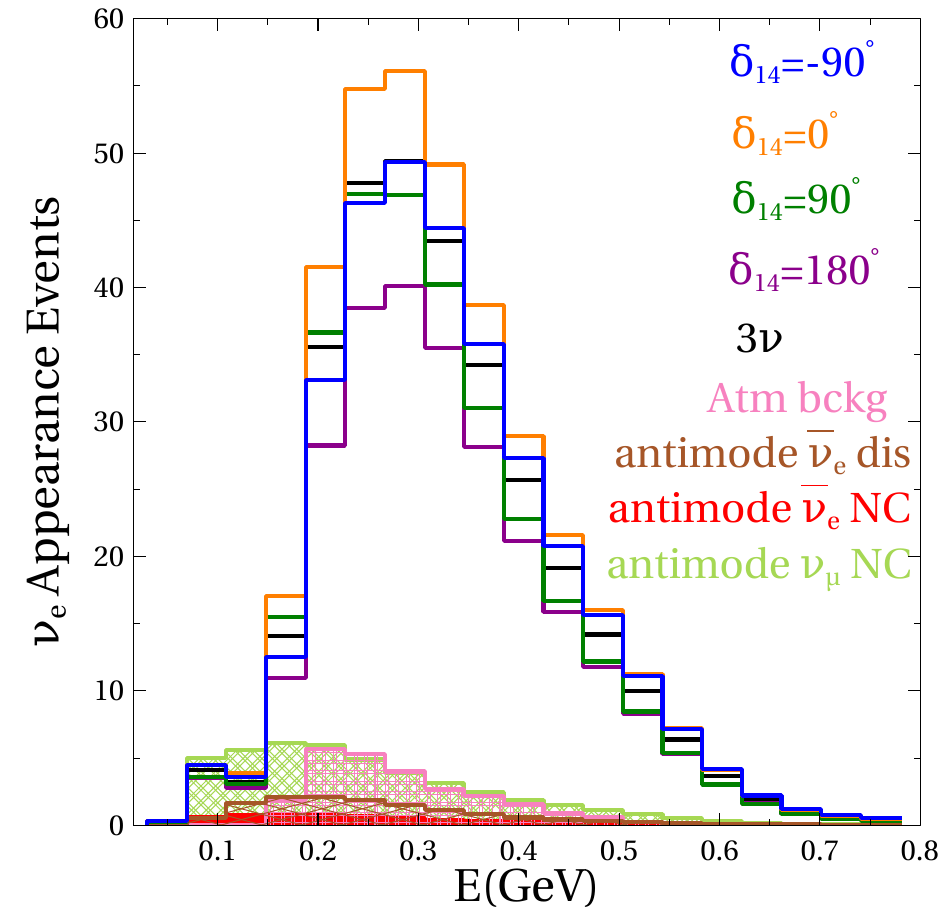}
\includegraphics[scale=0.45]{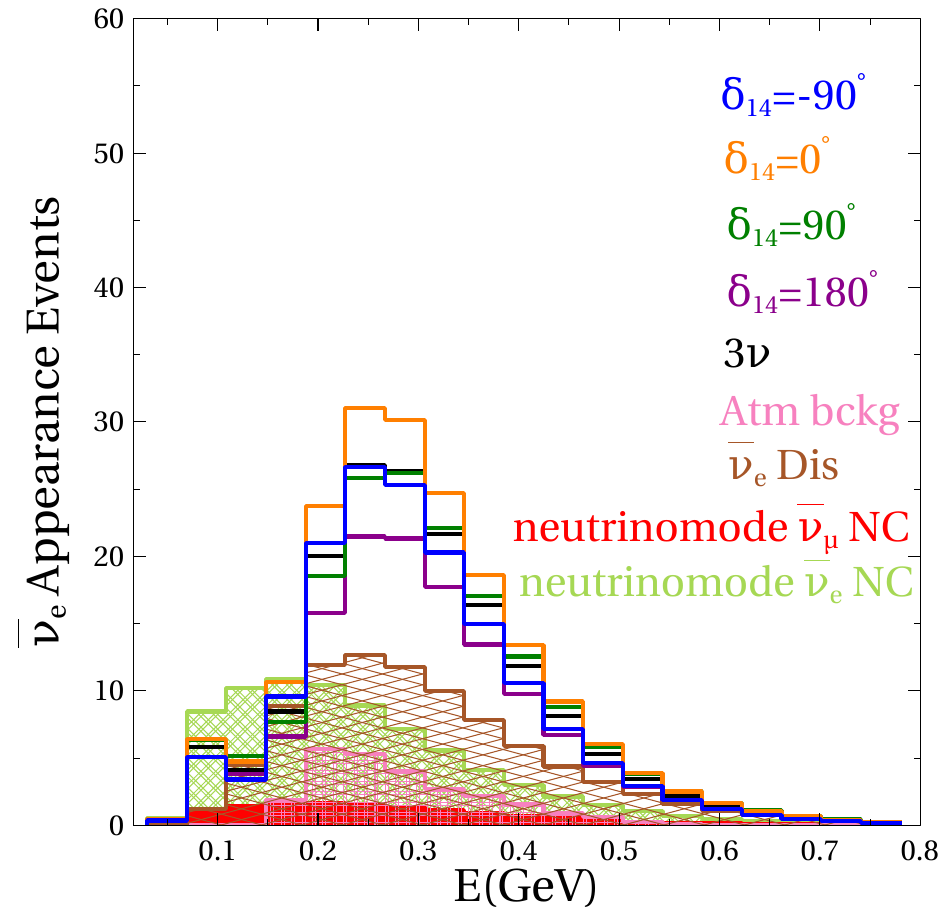}
\caption{The expected number of signal events are plotted against the reconstructed neutrino energy. The black curve refers to 3 flavor case while the colored histograms are for (3+1) scheme with different values of $\delta_{14}$ as mentioned in the legend. The left panel corresponds to $\nu_e$ appearance events while the right one is for $\overline{\nu}_e$ appearance events.}
\label{event}
\end{figure*} 
From the existing literature, we know that the number of events in the i-th energy bin are given by
\begin{equation}
N_i=\frac{Tn_n \epsilon}{4\pi L^2}\int_0^{E_{max}}dE \int_{E^{min}_{A_i}}^{E^{max}_{A_i}}dE_A \phi (E) \sigma_{\nu_e}(E)R(E,E_A)P_{\mu e}(E)
\end{equation}
where T is the total running time, $n_n$ is the number of target nucleons in the detector, $\epsilon$ is the detector efficiency, $\phi (E)$ is the neutrino flux, $\sigma_{\nu_e}$ is the neutrino interaction cross-section and R(E,$E_A$) is the $Gau\beta ian$ energy resolution function of detector. The quantities E and $E_A$ are the true and reconstructed (anti-)neutrino energies respectively and L is the baseline.

The numerical results have been performed using the GLoBES software~\cite{Huber:2004ka,Huber:2007ji} and its additional plugin~\cite{Kopp:2006wp,Kopp:2007ne} required to incorporate the new physics arising from the presence of sterile neutrino. We plot number of events against the reconstructed energy. It is found that the maximum flux peaks about 0.3 GeV, so we have maximum number of events in that energy bin. In the fig~\ref{event}, the the thick black lines depicts the 3 flavor scenario. The other colored histograms are drawn in the 3+1 scenario  for the different values of $\delta_{14}$. The number of $\nu_e$ appearance events are almost double of the number of events of $\overline{\nu}_e$ events since the cross section of neutrino interaction is very much different from the anti-neutrino interaction cross-section. For $\nu_e$ appearance events, we have shown the four possible backgrounds, where a pink color histogram represents background coming from atmospheric neutrinos while the brown color histogram  is for $\overline{\nu}_e$ disappearance events in anti neutrino mode run of $\rm{MOMENT}$ experiment. Additionally, the red and light green color hisogram represents the background coming from $\overline{\nu}_e$ and $\nu_\mu$ neutral currents, respectively. Similarly, for $\overline{\nu}_e$ appearance events, the background mainly comes from charge misidentification, atmospheric neutrinos, and neutral current backgrounds. It clearly emerges that the background effect is more prominent in $\overline{\nu}_e$ appearance events compared to $\nu_e$ appearance events.

\section{CP violation sensitivities in presence of sterile neutrino}
\label{sec:CP}
\subsection{Chi-square analysis}
 \label{sec:Chi}
The $\chi^2$ square analysis is performed to determine the sensitivity of an experiment in finding out the precise value of oscillation parameters. It is performed by comparing the simulated true event rates from the present best fit data with the events generated by the test hypothesis. Also, the theoretical uncertainties and the systematic errors at the experimental level are incorporated using the method of pulls in calculating $\chi^2$ as~\cite{Huber:2002mx,Fogli:2002pt},
\begin{eqnarray}
\chi^2 (p_{\rm true},p_{\rm test}) = \min_{\xi}\big[\sum_{i\,\in\,\mbox{bins}}\frac{(N^{\rm true}_i - N^{\rm test}_i(\xi))^2}{N^{\rm true}_i}+ \frac{\xi^2}{\sigma^2_{\xi}}\big]
\end{eqnarray}
 \noindent
where the nuisance parameter is denoted by $\xi$ and the corresponding systematic error is presented by $\sigma_{\xi}$. The terms involving the nuisance parameters are called pull terms. In order to counter the effect of systematic errors, the penalty term $\xi^2$ is added. The nuisance parameters are dependent on the fiducial mass of the detector used in a particular experiment as well as on the other experimental properties like flux normalization and cross section. The minimisation of $\chi^2$ is obtained by marginalizing the oscillation parameter space. Therefore, one add another penalty terms called priors to $\chi^2$. The mathematical expression for the prior is given by
\begin{eqnarray}
\chi^2_{\rm prior} = \bigg(\frac{N^{\rm true}- N^{\rm test}}{\sigma^{\rm true}}\bigg)^2
\end{eqnarray}
\noindent
 In our analysis, we looked at the sensitivity of MOMENT experiment to determine the precise value of CP phase in the presence of sterile neutrino. In the three flavor scenario as seen by eq.\,\eqref{P_vac}, the CP violations are induced by the presence of $\sin\delta_{13}$ term. The $\chi^2$ analysis is carried out with the statistical significance at which we can reject no CP violation test hypothesis as,
\begin{eqnarray}
\chi^2 = \frac{(N(\delta^{\rm tr}_{\rm CP})-N(\delta^{\rm test}_{\rm CP}=0,180))^2}{N(\delta^{\rm tr}_{\rm CP})}
\end{eqnarray}
\noindent
We have fixed the mixing angles $\theta_{12}$, $\theta_{13}$ and $\theta_{14}$ to their best fit values as mentioned in table~\ref{table_1} in both true and test spectrum. We look at the marginalization over the parameters $\theta_{23}$ and the mass-squared difference $\Delta m^2_{31}$ in two different ways as follows
\begin{enumerate}
\item First Case: $\Delta m^2_{31}$ is kept fixed for both 3 flavor and 3+1 flavor scheme while $\theta_{23}$ is marginalized over the range mentioned in table~\ref{table_1}.
\item Second Case: Both $\Delta m^2_{31}$ and $\theta_{23}$ are marginalized in 3 and 3+1 flavor scheme. Also the projected information on $\Delta m^2_{31}$ is added in the form of priors as 
\begin{eqnarray}
\chi^2_{\rm prior} = \bigg(\frac{\Delta m^2_{31}(\rm true)-\Delta m^2_{31}(\rm test)}{\sigma(\Delta m^2_{31})}\bigg)^2
\end{eqnarray}
where $\sigma(\Delta m^2_{31})$ is the $1\sigma$ error on $\Delta m^2_{31}$.
\end{enumerate}

\begin{figure*}[h]
\includegraphics[scale=0.55]{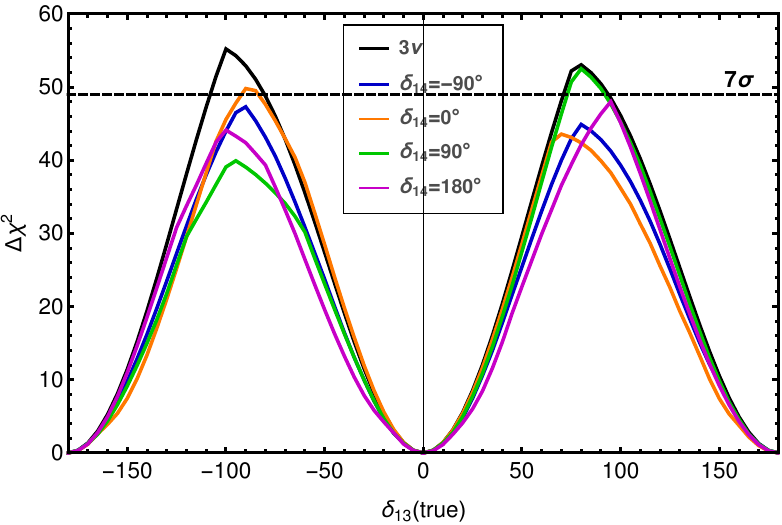}
\includegraphics[scale=0.4250]{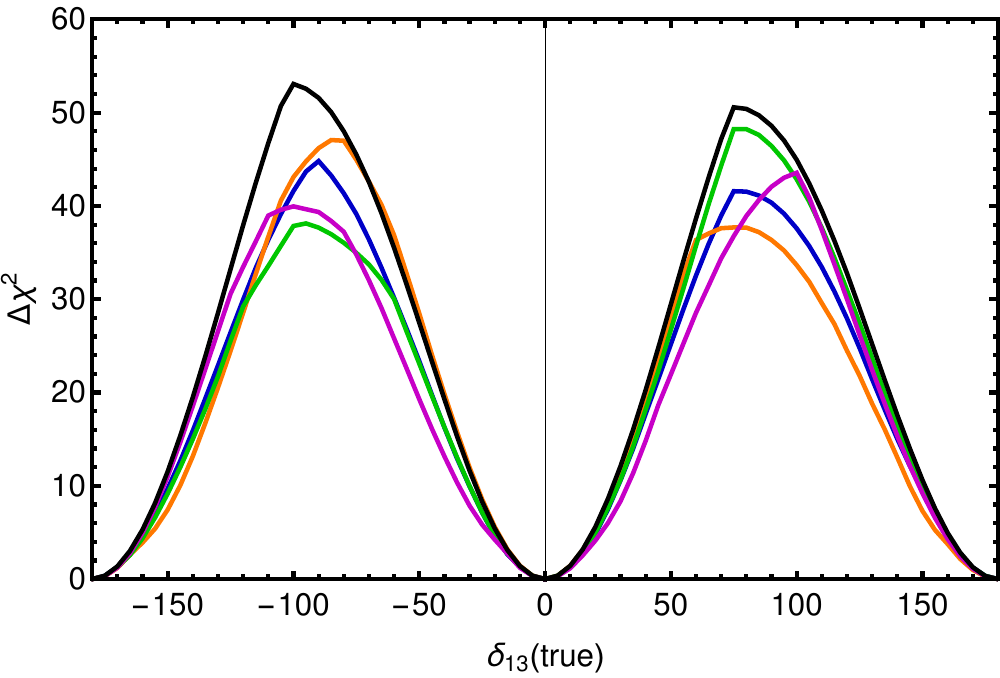}
\caption{The figure indicated the potential of MOMENT experiment for the discovery of CP violation induced by the fundamental phase $\delta_{13}$. The black curve shows the behavior under 3 flavor scheme while the colored curves in each panel are of different values of $\delta_{14}$. The left figure $\Delta m^2_{31}$ is kept fixed for both 3 and 3+1 flavor mixing while we marginalize over $\theta_{23}$ value whereas in the right figure, we marginalised over both hierarchy $\Delta m^2_{31}$ and $\theta_{23}$.}
\label{fig cp}
\end{figure*}

The numerical results are displayed in figure~\ref{fig cp}. The left-panel of the figure corresponds to the fixed value of $\Delta m^2_{31}$ and varying $\theta_{23}$ while right-panel of the figure corresponds to the second case of marginalizing over  both the mass hierarchy $\Delta m^2_{31}$ and $\theta_{23}$. The solid black solid line represents the estimated value of $\Delta \chi^2$ for three flavor scenario while all other color lines depict the analysis for 3+1 flavor scenario. While performing the $\chi^2$ analysis for discovery potential of CP phases we have considered equal neutrino and anti-neutrino mode for a run time of (5+5) years. In each case we have considered four different values of true $\delta_{14}$ phase as considered in probability analysis while its test value is marginalized. The figure clearly depicts that CP sensitivities decreases in the presence of sterile neutrino, primarily because of the degeneracies between the fundamental and active-sterile CP phase.

\subsection{Reconstructing the CP phases}
\begin{figure*}[tbh]
	 \includegraphics[scale=0.54]{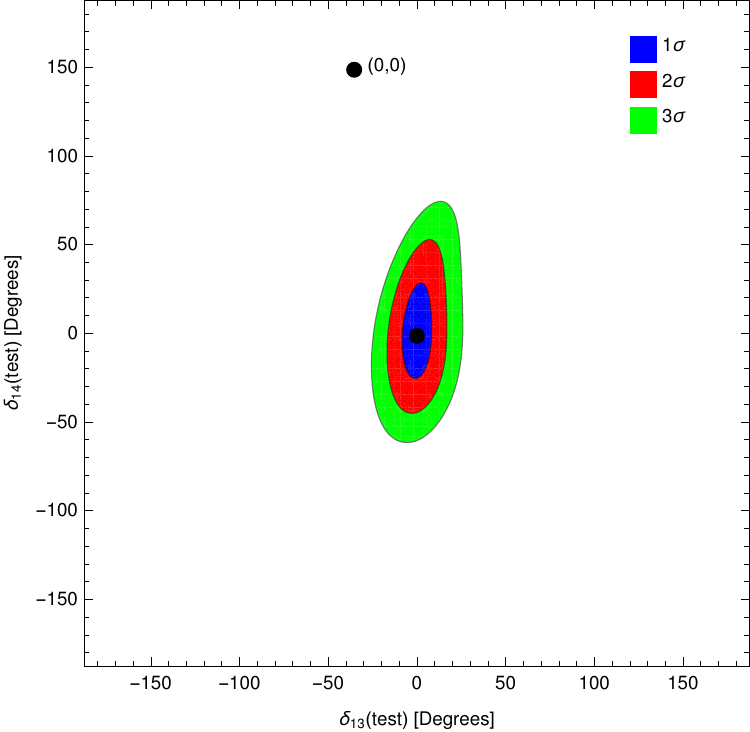}
     \includegraphics[scale=0.53]{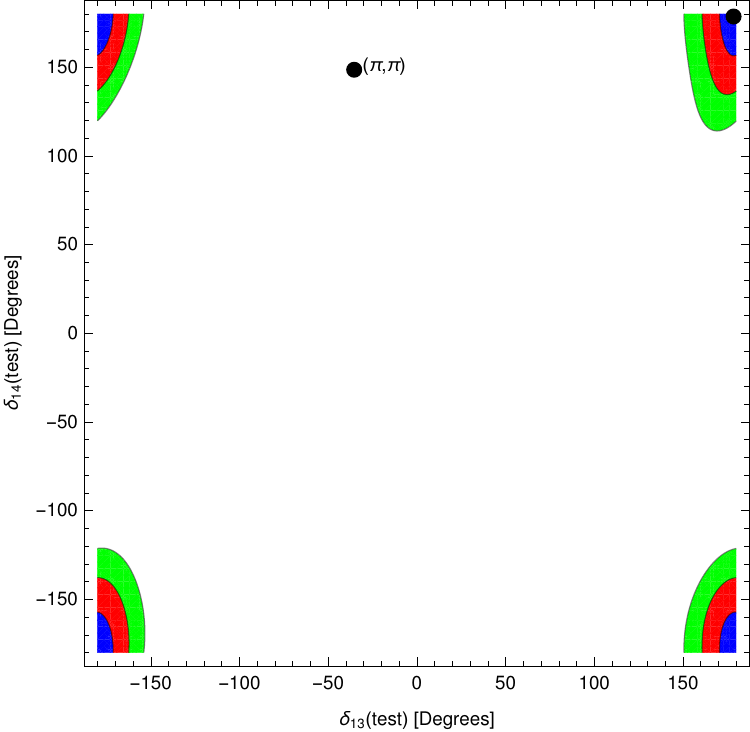} 
	 \includegraphics[scale=0.55]{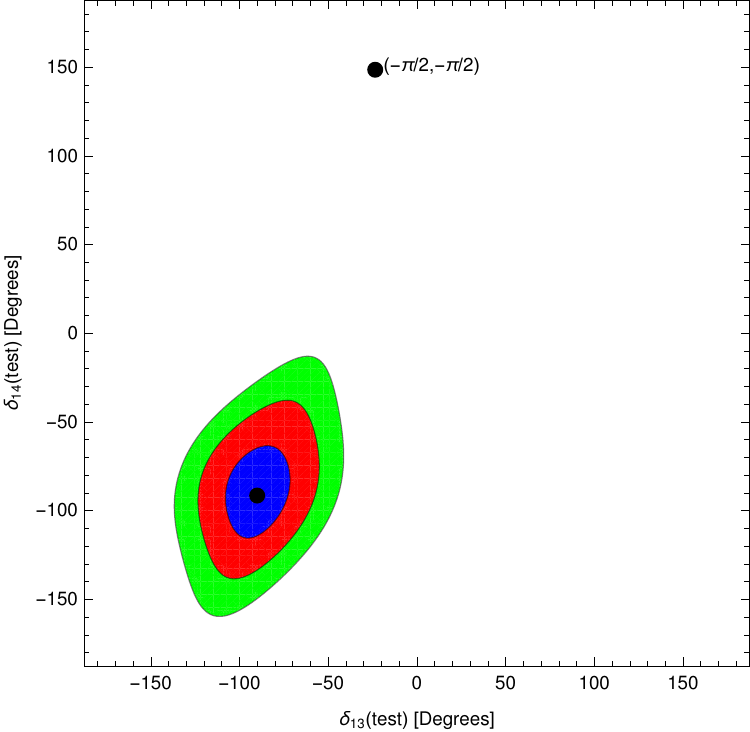}
     \includegraphics[scale=0.55]{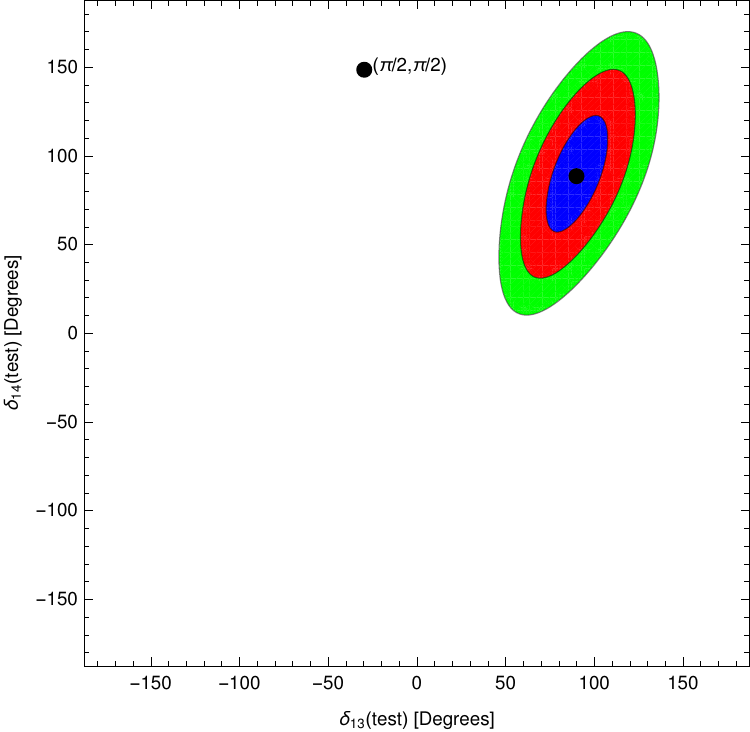}
	\caption{ Contour plots in the plane of $\delta_{13}$ (test) and $\delta_{14}$ (test) for four different values of CP phases $\delta_{13}$ and $\delta_{14}$. The discovery potential of $\delta_{13}$ in the contour plot of $\delta_{13}$ vs $\delta_{14}$ plane has been presented by blue, red, and green band with $1\sigma$, $2\sigma$, and $3\sigma$ confidence level respectively. The left(right) panel of first row corresponds to (0,0) and ($\pi,\pi$) respectively while the left(right) panel of second row corresponds to ($-\pi /2,-\pi /2$) and ($\pi /2,\pi /2$)
	} 
	\label{fig:recoCP}
\end{figure*}

In the last subsection, we looked at the CP-violation discovery potential of the MOMENT experiment by performing $\chi^2$ analysis.  But there is an another way to extract the complementary information by reconstructing the values of two CP phases $\delta_{13}$ and \blue{$\delta_{14}$}, independent of the amount of CP violation. We look at the contour plots by reconstructing the CP phases in $\delta_{13}-\delta_{14}$ plane. The test values of both the CP phases are varied from $(-\pi, \pi)$ and contours are shown at one, two, and three sigma level, simultaneously. The four plots represents the regions reconstructed for the four benchmark values considered in figure~\ref{fig:recoCP}. The solid black dot in the upper row represents CP-conserving scenarios with values (0,0), $(\pi,\pi)$ while the bottom panel has CP-violating picture with $(-\pi /2, -\pi /2)$, $(\pi /2, \pi /2)$. 
The uncertainties in the standard CP Phase $\delta_{13}$ (sterile CP phase $\delta_{14}$) can be traced by the horizontal (vertical) spanning of resulting contours. The contours are narrower along the test $\delta_{13}$ while they are broader along the test $\delta_{14}$ axis. Thus, reconstruction of $\delta_{13}$ is better than $\delta_{14}$, one can understand it by the fact that mixing angle $\theta_{13}$ is known with much precisely than mixing angle $\theta_{14}$. To have a quantitative idea of reconstructing the true values of CP phases, we calculate the horizontal and vertica span of contours at $3\sigma$ level for $\delta_{13}$ and $\delta_{14}$ respectively. The maximum range of uncertainities in reconstructing the true values of CP phases are mentioned in Table~\ref{table_3}. Our results are predicts remarkable sensitivity for determining the precise value of the CP phase $\delta_{13}$ in the presence of sterile neutrino.

\begin{table}[h!] 
	\centering
	\setlength{\extrarowheight}{0.1cm}
	\begin{tabular}{|c|c|c|}
		\hline
      \textbf{CP Phases} & \textbf{Values to be reconstructed} &\textbf{Reconstructed Range } \\
      \cline{2-3}
      \hline
      \hline
      ($\delta_{13}$,$\delta_{14}$) &  ($0^\circ$,$0^\circ$) &  $-30^\circ \leq \delta_{13} \leq 30^\circ $ \,\,\,\ $-30^\circ\leq \delta_{14} \leq 80^\circ$
      \\
      \hline
      ($\delta_{13}$,$\delta_{14}$) &  ($180^\circ$, $180^\circ$) &$150^\circ \leq \delta_{13} \leq 180^\circ $ \,\,\,\ $110^\circ\leq \delta_{14} \leq 180^\circ$ \\
      \hline
      ($\delta_{13}$,$\delta_{14}$) & ($-90^\circ$,$-90^\circ$) &$-140^\circ \leq \delta_{13} \leq -45^\circ$ \,\, $-165^\circ\leq \delta_{14} \leq -15^\circ$ \\
      \hline
      ($\delta_{13}$,$\delta_{14}$) & ($90^\circ$, $90^\circ$) & $45^\circ \leq \delta_{13} \leq 140^\circ $  \,\,\ $10^\circ\leq \delta_{14} \leq 170^\circ$  \\
          \hline
	\end{tabular}
	\caption{The reconstrcuted range of test values of $\delta_{14}$, $\delta_{14}$ for the four choices of their true values at $3\sigma$ level.}
	\label{table_3}	
	\end{table} 
\cleardoublepage
\section{Conclusions and Outlook}
\label{sec:Conclusions}
We have addressed the physics potential of MOMENT experiment to emphasis the role of sterile neutrino presence over the standard three flavor oscillation parameters. We have shown the transition probabilities for both neutrinos and anti-neutrinos in 3+1 flavor scheme and looked at the space spanned by the CP trajectory curves. The performance of MOMENT in understanding the CP violation sensitivities induced by the fundamental CP phase $\delta_{13}$ and new CP phase arising from active-sterile mixing has been explored. We found that loss of CP sensitivity is dependent on the different values of $\delta_{14}$ phase. The discovery potential of CP violation in 3 flavor scheme is quite significiant at $7\,\sigma$ level though it gets reduced in the presence of unknown CP phase $\delta_{14}$. The reduction might has arose from the degeneracies among the two CP phases. We have also assessed the capability of MOMENT experiment in reconstructing the true values of such CP-phases.

\subsubsection*{Acknowledgments}
Kiran Sharma would like to thank Ministry of Education for the financial support for carrying out this research work. KS is very thankful to Dr. Sabya Sachi Chatterjee for the fruitful discussion carried time to time for the betterment of this work.

\section{APPENDIX}
\appendix

\section{Detailed Description of Biprobability analysis.}
 \label{appendix}
The analytic expression for the neutrino transition probabilities in case of $3+1$ neutrino oscillation are recasted into a compact form as,
\begin{eqnarray}
 P(\nu) \equiv P = P_0 + A \cos \big( \Delta + \delta_{13} \big) 
           + B \sin \big( \Delta - \delta_{14}+ \delta_{13} \big)  \,
 \label{eq:Parametric+nu}
 \end{eqnarray}
where the first term $P_0= P^{\rm {ATM}} \!\! \simeq\,  4 s_{23}^2 s^2_{13}  \sin^2{\Delta}$ is independent of phase factor while the factors which are independent of phase contained in second and third term are, $A=8 s_{13} s_{12} c_{12} s_{23} c_{23} (\alpha \Delta)\sin \Delta$ and $B=4 s_{14} s_{24} s_{13} s_{23} \sin\Delta$. 
After few simplification using trigonometric relations, the transition probability given in eq.(\ref{eq:Parametric+nu}) modifies to
\begin{eqnarray}
 P = P_0 + A^\prime \cos\delta_{13} 
           + B^\prime \sin \delta_{13} 
 \label{eq:Parametric+nu+modified}
 \end{eqnarray}
 
 Similarly, the simplified antineutrino transition probability is given by
 \begin{eqnarray}
 \overline{P} = \overline{P}_0 + \overline{A}^\prime \cos\delta_{13} 
           - \overline{B}^\prime \sin \delta_{13} 
 \label{eq:Parametric+nu+modified}
 \end{eqnarray}
 
 where the coefficients $A^\prime$, $B^\prime$,$\overline{A}^\prime$ and $\overline{B}^\prime$ are defined as follows
 \begin{equation}
 A^\prime = A \cos \Delta + B \sin \big(\Delta-\delta_{14} \big)
 \label{A prime}
 \end{equation}
 
 \begin{equation}
 B^\prime = -A \sin \Delta + B \cos \big(\Delta-\delta_{14} \big)
 \label{B prime}
 \end{equation}
 
 \begin{equation}
 \overline{A}^\prime = \overline{A} \cos \Delta + \overline{B} \sin \big(\Delta+\delta_{14} \big)
 \label{A_bar prime}
 \end{equation}
 
 \begin{equation}
 \overline{B}^\prime = -\overline{A} \sin \Delta + \overline{B} \cos \big(\Delta+\delta_{14} \big)
 \label{B_bar prime}
 \end{equation}
 
 The factors $\overline{A}^\prime$ and $\overline{B}^\prime$ are obtained from known relations of $A^\prime$ and $B^\prime$ by replacing $\delta_{14} \to -\delta_{14}$. Eliminating $\delta_{13}$ from the modified neutrino and antineutrino transition probabilities, we obtain,
  \begin{eqnarray}
\frac{1}{C^2}
   \bigg(\frac{P-P_0}{B^\prime} + \frac{\overline{P}-\overline{P}_0}{\overline{B}^\prime} \bigg)^2 
   + 
 \frac{1}{D^2}
   \bigg(\frac{P-P_0}{A^\prime} - \frac{\overline{P}-\overline{P}_0}{\overline{A}^\prime} \bigg)^2   = 1 
 \label{eq:elipse}
 \end{eqnarray}
where 
$C = {\frac{A^\prime}{B^\prime}+\frac{\overline{A}^\prime}{\overline{B}^\prime}}$ and $D = \frac{B^\prime}{A^\prime}+\frac{\overline{B}^\prime}{\overline{A}^\prime}$.
 \newline
  After few simplifications, we can effectively express the above equation as,
 \begin{eqnarray}
 && \bigg[\frac{1}{{B^\prime}^2 C^2} +  \frac{1}{{A^\prime}^2D^2}    
          \bigg]P^2 
            \nonumber \\ &&
 \hspace*{1cm}+   \bigg[\frac{2}{{B^\prime}{\overline{B}^\prime} C^2} -\frac{2}{{A^\prime}{\overline{A}^\prime} D^2}\bigg]P \overline{P}
     \nonumber \\ &&
 \hspace*{1cm}
 +  \bigg[\frac{1}{{\overline{B}^\prime}^2 C^2} +  \frac{1}
          {{\overline{A}^\prime}^2 D^2}\bigg]\overline{P}^2 
  \nonumber \\ &&
 \hspace*{1cm}
 + \bigg[\frac{-2P_0}{{B^\prime}^2 C ^2} + \frac{-2P_0}{{A^\prime}^2 D^2} +  \frac{-2\overline{P}_0}{{B^\prime}{\overline{B}}^\prime C^2} + \frac{-2\overline{P}_0}{{A^\prime}{\overline{A}^\prime}D^2}\bigg]P 
    \nonumber \\ &&
 \hspace*{1cm}
   + \bigg[\frac{-2\overline{P}_0}{{\overline{B}^\prime}^2 C^2} + \frac{-2\overline{P}_0}{{\overline{A}^\prime}^2 D^2} +\frac{-2P_0}{{B^\prime}{\overline{B}}^\prime C^2} + \frac{-2P_0}{{A^\prime}{\overline{A}^\prime} D^2}\bigg]\overline{P}   
    \nonumber \\ &&
 \hspace*{1cm}
 +     %
  \bigg[\bigg(\frac{1}{{B^\prime}^2 C^2} 
 +  \frac{1}{{A^\prime}^2 D^2}\bigg)P_0^2 +   \bigg(\frac{1}{{\overline{B}^\prime}^2 C^2} +  \frac{1}{{\overline{A}^\prime}^2 D^2}\bigg)\overline{P_0}^2      \nonumber \\ &&
 \hspace*{4cm}
 + \bigg(\frac{2}{{B^\prime}{\overline{B}^\prime} C^2}  
  - \frac{2}{{A^\prime}{\overline{A}^\prime} D^2}\bigg)P_0 \overline{P_0} -1\bigg]  = 0 
\label{ellipse_eqn}
 \end{eqnarray}
Comparing it with the general quadratic curve representing the equation of ellipse  
\begin{equation}
ax^2 +bxy + cy^2 +dx+ ey+f =0 
\end{equation}
where the counterclockwise angle of rotation from the x-axis to the major axis of the ellipse is defined by $\tan 2\theta = \frac{b}{a -c}$, we can understand the inclination of ellipse in the biprobability plots. 
\begin{eqnarray}
&& a = \frac{1}{{B^\prime}^2 C^2} +  \frac{1}{{A^\prime}^2 D^2} \nonumber \\ 
&& b = \frac{2}{{B^\prime}{\overline{B}^\prime} C^2} -\frac{2}{{A^\prime}{\overline{A}^\prime} D^2} \nonumber \\
&& c = \frac{1}{{\overline{B}^\prime}^2 C^2} +  \frac{1}{{\overline{A}^\prime}^2 D^2}
\end{eqnarray}
Simplifying the results under the assumption that matter effects induce negligible perturbations on the interference terms (i.e. $A = \overline{A}$, $B = \overline{B}$) and using equations ~\eqref{A prime}, \eqref{B prime}, ~\eqref{A_bar prime} and ~\eqref{B_bar prime}, the angle of inclination becomes
\begin{eqnarray}
\tan 2\theta = \frac{(B^2 - A^2)\cos 2\Delta - 2AB\sin 2\Delta \cos\delta_{14}}{2 AB\sin\delta_{14}}
\end{eqnarray}

\bibliographystyle{JHEP}
\bibliography{sterile}

\end{document}